\documentclass[%
 reprint,
 amsmath,amssymb,
 aps,
]{revtex4-1}

\usepackage{graphicx}% Include figure files
\usepackage{dcolumn}% Align table columns on decimal point
\usepackage{bm}% bold math
\usepackage{comment}
\usepackage{braket}

\usepackage{color}

\begin{document}

\preprint{APS/123-QED}

\title{Decay properties of Roper resonance in the holographic QCD}

\author{Daisuke Fujii}
 \email{daisuke@rcnp.osaka-u.ac.jp}
\author{Atsushi Hosaka}%
 \altaffiliation[Also at ]{Advanced Science Research Center, Japan Atomic Energy Agency, Tokai, Ibaraki 319-1195 Japan.}
  \email{hosaka@rcnp.osaka-u.ac.jp}
\affiliation{%
 Research Center for Nuclear Physics(RCNP), Osaka University, Ibaraki 567-0048, Japan.\\
}%

\date{\today}% It is always \today, today,

\begin{abstract}
We investigate the one pion decay of the Roper resonance 
$N^*(1440) \to N \pi$ in the Sakai-Sugimoto model of the holographic QCD.  
The nucleon and Roper resonance emerge as ground and first excited states of the collective radial motion of the instanton in the four dimensional space with one extra dimension.  
It is found that the ratio of the $\pi NN^*$ and $\pi NN$ couplings, and hence the ratio of $g_A^{NN^*}$ and $g_A^{NN}$, is well reproduced in comparison with the experimental data.  
The mechanism of this result is due to the collective nature of excitations, 
which is very different from that of the single particle nature  of the constituent quark model.  
Our results are obtained in the large-$N_c$ and large $\lambda$ ('t~Hooft coupling) limit 
which are useful to test 
how baryon resonances share what are expected in these limits.  
\end{abstract}

\maketitle

%====================================================
\section{\label{Introduction}Introduction}
%====================================================

The Roper resonance $N^*(1440)$ 
is the first excited state of the nucleon with the spin and parity $J^{P}=1/2^{+}$~\cite{Roper:1964zza}.  
Its mass smaller than the negative parity nucleon $N(1535)$ 
has attracted great amount of interests because the naive quark model predicts
the mass of the Roper resonance much higher than that of the negative parity state.   
To resolve this problem, and also to reproduce the electromagnetic transitions, 
the importance of the meson cloud has been emphasized~\cite{Suzuki:2009nj,Burkert:2017djo}.
Turning to strong decays, an almost vanishing partial decay width 
of one pion emission when computed by the leading order terms
of non-relativistic expansion of the pion-quark interaction 
disagree with the large value of the experimental data.  
While it has been pointed out recently that higher order corrections 
can improve this significantly~\cite{Arifi:2021orx}, 
this problem should be further investigated.  

The relatively low mass has lead to the idea of collective vibrational mode along the radial direction~\cite{Brown:1983ib}.  
Extensive discussions were made in the Skyrme model in 1980's, 
where the soliton's radial vibrations were investigated in various context~\cite{Kaulfuss:1985na,Mattis:1984ak,Walliser:1984wn,Lee:1988yz}.  

Later the solitonic picture of baryons has been further strengthened 
by the holographic QCD.
The Sakai-Sugimoto model is one of successful descriptions of hadrons in the holographic QCD 
based on the D4-D8 brane construction~\cite{Sakai:2004cn,Sakai:2005yt}.  
They have derived an effective action of the flavor gauge field 
in the five dimensional space (four space-time and one extra dimension), 
implementing  the spontaneous breaking of chiral symmetry 
leading to the successful low-energy effective action of hadrons. 
Moreover the extra dimension of the model naturally accommodates 
various excited states of hadrons. 

In the holographic model, baryons emerge as instantons
of the five-dimensional space~\cite{Hata:2007mb}, which is 
very much the same as the Skyrme mode, baryons as chiral solitons~\cite{Skyrme:1962vh,Adkins:1983ya}.  
Such a baryon structure looks very different from the one of the quark model. 
Baryon dynamics is dominated by the collective motions of instantons/solitons, 
while that of the quark model by single-particle motions of quarks.  
Interestingly, it was found that the resulting Roper and the negative parity resonance~\cite{Hata:2007mb}
are degenerate and appear very close to the observed masses.  
This is one of good features of the holographic QCD for baryons.  

The holographic baryons have been further studied by Hata et al~\cite{Hata:2008xc} and 
by Hashimoto et al~\cite{Hashimoto:2008zw} for various static properties 
of the nucleon including electromagnetic and weak coupling constants.
Inspired by these works, we would like to further study 
the properties of the Roper resonance in the holographic model.  
In this paper, we investigate  the one pion emission decay.
It is the axial transition between the Roper resonance and the nucleon, and is dictated by the 
transition matrix element of the axial current.  
Following Ref~\cite{Hashimoto:2008zw}, 
we define chiral currents by introducing the external gauge field that couples to the currents. 
By calculating the matrix elements of the obtained axial current, 
the axial coupling and hence decay width are calculated.
The results are compared with the experimental data.
The model and computation procedures are realized in the large-$N_c$ 
and large 't~Hooft coupling $\lambda$ limits.
Hence our study provides the measure to what extend hadron properties
share the features of these limits.   

This paper is organized as follows. 
In section \ref{axial current}, we present the actions used in this paper and derive the solutions of the equations of motion.
Then we define the chiral currents and obtain their concrete expressions by the solutions.  
In section \ref{decay properties}, we compute matrix elements of the axial currents for the nucleon 
and that of the Roper to the nucleon transitions.  
The resulting decay width is compared with the experimental data.
We discuss the ratio of $g^{NN^{*}}_{A}$ and $g^{NN}_{A}$, 
and compare with the data  carefully.  
Final section \ref{conclusion} is for some discussions and summary of  the present work.

%=========================================
\section{\label{axial current}Axial current}
%=========================================

%=========================================
\subsection{\label{classical solutions}Classical solutions and collective quantization}
%=========================================

Let us start by briefly summarizing how the baryon states are obtained in the Sakai-Sugimoto model 
by collectively quantizing the instanton solution.  
The action of hadron effective theory is composed of 
the Yang-Mills term $S_{YM}$ and the Chern-Simons term $S_{CS}$, 
\begin{eqnarray}
S=S_{YM}+S_{CS}
\end{eqnarray}
where 
\begin{eqnarray}
S_{YM}&&=-\kappa\int d^{4}xdz{\rm tr}\left[\frac{1}{2}h(z)\mathcal{F}^{2}_{\mu\nu}+k(z)\mathcal{F}^{2}_{\mu z}\right], \notag \\
S_{CS}&&=\frac{N_{c}}{24\pi^{2}}\int_{M^{4}\times \mathbb{R}}\omega_{5}(\mathcal{A}), \notag \\
\kappa&&=\frac{\lambda N_{c}}{216\pi^{3}}=a\lambda N_{c}.
\end{eqnarray}
In these equations $N_{c}$ is the number of colors, $\lambda$ the 't~Hooft coupling, 
and the indices $\mu,\nu=0,1,2,3$ are for the 4-dimensional space-time.
The curvatures along the extra dimension $z$ are defined by 
\begin{eqnarray}
h(z)=(1+z^{2})^{-1/3}, \ \  k(z)=1+z^{2}.
\end{eqnarray}
The 1-form $\mathcal{A}$ expresses
$\mathcal{A}=A_{\alpha}dx^{\alpha}+\hat{A}_{\alpha}dx^{\alpha}$ 
which consists of the flavor SU(2) part $A_{\alpha}$ and the U(1) part $\hat{A}_{\alpha}$ with $\alpha=0,1,2,3,z$. 
The Chern-Simons 5-form is given by
\begin{eqnarray}
\omega_{5}(\mathcal{A})={\rm tr}\left(\mathcal{A}\mathcal{F}^{2}-\frac{i}{2}\mathcal{A}^{3}\mathcal{F}-\frac{1}{10}\mathcal{A}^{5}\right).
\end{eqnarray}

In general, it is difficult to analytically solve the equations of motion
in the presence of the curvatures $h(z)$ and $k(z)$.  
However, it can be simplified in the large $\lambda$ limit 
since the instanton profile is 
localized around $z\sim 0$ as proportional to $\lambda^{-1/2}$, 
where we can set $h(z) = k(z) = 1$.  
Therefore, the following instanton solution is available, with $M = 1,2,3, z$, 
\begin{eqnarray}
&&A^{cl}_{M}\left(\mathbf{x},z\right)=-if\left(\xi\right)g\partial_{M}g^{-1} \label{A^cl_M}, \notag \\
&&A^{cl}_{0}=0, \\
&&\hat{A}^{cl}_{M}=0, \notag \\
&&\hat{A}^{cl}_{0}=\frac{1}{8\pi^{2}a}\frac{1}{\xi^{2}}\left[1-\frac{\rho^{4}}{\left(\xi^{2}+\rho^{2}\right)^{2}}\right],
\end{eqnarray}
where
\begin{eqnarray}
&&g\left(\mathbf{x},z\right)=\frac{\left(z-Z\right)-i\left(\mathbf{x}-\mathbf{X}\right)\cdot\boldsymbol{\tau}}{\xi}, \notag
\end{eqnarray}
with $\left({\bm X}, Z \right)$ and $\rho$  the location and size of the instanton, respectively.
The profile function $f\left(\xi\right)$ is  given by
\begin{eqnarray}
&&f\left(\xi\right)=\xi^{2}/\left(\xi^{2}+\rho^{2}\right), \notag \\
&&\xi=\sqrt{\left(\mathbf{x}-\mathbf{X}\right)^{2}+\left(z-Z\right)^{2}}. \notag
\end{eqnarray}

The classical instanton solution needs to be quantized for the physical nucleon and Roper resonances. 
This can be done by the collective coordinate method, where the relevant time dependent dynamical variables are, 
${\bm X}, Z$, $\rho$ and the rotational variable in the isospin and spin space.  
As shown in Ref.~\cite{Hata:2007mb} the time dependent gauge field is given by
\begin{eqnarray}
&&A_{M}\left(t,x^{N}\right) \notag \\
&&=VA^{cl}_{M}\left(x^{N};X^{N}\left(t\right),\rho\left(t\right)\right)V^{-1}-iV\partial_{M}V^{-1}, \\
&&\Phi\left(t,x^{M}\right) =-iV^{-1}\dot{V} \nonumber \\
&&= -\dot{X}^{M}\left(t\right)A^{cl}_{M}+\chi^{a}f\left(\xi\right)g\frac{\tau^{a}}{2}g^{-1}, \notag \\
&&\chi^{a}=-i{\rm tr}\left(\tau^{a}{\bm a}^{-1}\dot{{\bm a}}\right), \notag
\end{eqnarray}
where ${\bm a}=a_{4}+ia_{a}\tau^{a}$ is for the spin and isospin rotation.  
By using this gauge field, 
we find the collective Hamiltonian as
\begin{eqnarray}
&&H=-\frac{1}{2M_{0}}\left(\partial_{\vec{X}}^{2}+\partial_{Z}^{2}\right)-\frac{1}{4M_{0}}\partial^{2}_{y^{I}}+U\left(\rho,Z\right), \notag \\
&&U\left(\rho,Z\right)=M_{0}+\frac{M_{0}}{6}\rho^{2}+\frac{N^{2}_{c}}{5M_{0}}\frac{1}{\rho^{2}}+\frac{M_{0}}{3}Z^{2},
\end{eqnarray} 
where $M_{0}=8\pi^{2}\kappa$ is the classical soliton mass~\cite{Hata:2007mb}, and $y_{I}$ is related to the 
orientation coordinates by 
$y_{I}=\rho a_{I}$.  
The baryon states are labeled by its momentum $\vec{p}$ and quantum numbers 
$(l,I_{3},s_{3},n_{\rho},n_{z})$, 
where $l/2$ is the equal isospin and spin values;
$I_{3},s_{3}$ are the third components of the isospin and spin; 
and $n_{\rho},n_{z}$ are the quanta for oscillations along the radial and $z$-directions.  
For the spin up proton ($I_3 = 1/2, \ s_{3}=1/2$) 
with a finite momentum $\vec{p}$, 
the wave functions of ground and Roper resonance are given as~\cite{Hata:2007mb,Hashimoto:2008zw}
\begin{eqnarray}
\psi_{N}\ &\propto& \ e^{i\vec{p}\cdot\vec{X}}R_{N}\left(\rho\right)\psi_{Z}\left(Z\right)\left(a_{1}+ia_{2}\right), 
\notag \\
\psi_{N^{*}(1440)}\ &\propto& \ e^{i\vec{p}\cdot\vec{X}}R_{N^{*}}\left(\rho\right)\psi_{Z}\left(Z\right)\left(a_{1}+ia_{2}\right), 
\label{wave func} 
\end{eqnarray}
where
\begin{eqnarray}
&&R_{N}\left(\rho\right)=\rho^{-1+2\sqrt{1+N^{2}_{c}/5}}e^{-\frac{M_{0}}{\sqrt{6}}\rho^{2}}, \notag \\
&&R_{N^{*}}\left(\rho\right)=\left(\frac{2M_{0}}{\sqrt{6}}\rho^{2}-1-2\sqrt{1+\frac{N^{2}_{c}}{5}}\right)R_{N}\left(\rho\right), \\
&&\psi_{Z}\left(Z\right)=e^{-\frac{M_{0}}{\sqrt{6}}Z^{2}}.
\end{eqnarray}
We note that the wave function for the $z$ oscillation is the lowest ($n_z = 0$) for both 
the nucleon and Roper resonance.  
Thus the only difference between them is in the radial part, 
$R_{N}(\rho)$ and $R_{N^*}(\rho)$.  

%=========================================
\subsection{\label{extended solutions}The asymptotic solution of the instanton}
%=========================================

The BPST instanton that we have summarized in the previous subsection is only an approximate solution 
in the large $\lambda$ limit where the instanton size is small. 
This can be used for the computation of baryon masses.  
However, for the computation of currents which are defined at $|z| \to \infty$
such a solution is not suited.  
As shown in Ref.~\cite{Hashimoto:2008zw} 
we need to find the solution that is properly extended to the  large $|z|$ region
to obtain the well-defined currents.  
In this paper, we simply summarize the final result of such a solution; 
\begin{widetext}
\begin{eqnarray}
&&\hat{A}_{0}=-\frac{1}{2a\lambda}G\left(\vec{x},z;\vec{X},Z\right), \notag \\
&&\hat{A}_{i}=\frac{1}{2a\lambda}\left[\dot{X}^{i}+\frac{\rho^{2}}{2}\left(\frac{\chi^{a}}{2}\left(\epsilon^{iaj}\frac{\partial}{\partial X^{j}}-\delta^{ia}\frac{\partial}{\partial Z}\right)+\frac{\dot{\rho}}{\rho}\frac{\partial}{\partial X^{i}}\right)\right]G\left(\vec{x},z;\vec{X},Z\right), \notag \\
&&\hat{A}_{z}=\frac{1}{2a\lambda}\left[\dot{Z}+\frac{\rho^{2}}{2}\left(\frac{\chi^{a}}{2}\frac{\partial}{\partial X^{a}}+\frac{\dot{\rho}}{\rho}\frac{\partial}{\partial Z}\right)\right]H\left(\vec{x},z;\vec{X},Z\right), \\
&&A_{0}=4\pi^{2}\rho^{2}i{\bm a}\dot{{\bm a}}^{-1}G\left(\vec{x},z;\vec{X},Z\right)+2\pi^{2}\rho^{2}{\bm a}\tau^{a}{\bm a}^{-1}\left(\dot{X}^{i}\left(\epsilon^{iaj}\frac{\partial}{\partial X^{j}}-\delta^{ia}\frac{\partial}{\partial Z}\right)+\dot{Z}\frac{\partial}{\partial X^{a}}\right)G\left(\vec{x},z;\vec{X},Z\right), \notag \\
&&A_{i}=-2\pi^{2}\rho^{2}{\bm a}\tau^{a}{\bm a}^{-1}\left(\epsilon^{iaj}\frac{\partial}{\partial X^{j}}-\delta^{ia}\frac{\partial}{\partial Z}\right)G\left(\vec{x},z;\vec{X},Z\right), \notag \\
&&A_{z}=-2\pi^{2}\rho^{2}{\bm a}\tau^{a}{\bm a}^{-1}\frac{\partial}{\partial X^{a}}H\left(\vec{x},z;\vec{X},Z\right).
\label{instanton3}
\end{eqnarray}
\end{widetext}
where the index $i$ runs $1 - 3$.   
In these equations, $G$ and $H$ are given by
\begin{eqnarray}
G\left(\vec{x},z;\vec{X},Z\right)=\kappa\sum^{\infty}_{n=1}\psi_{n}\left(z\right)\psi_{n}\left(Z\right)Y_{n}\left(|\vec{x}-\vec{X}|\right), \notag \\
H\left(\vec{x},z;\vec{X},Z\right)=\kappa\sum^{\infty}_{n=1}\phi_{n}\left(z\right)\phi_{n}\left(Z\right)Y_{n}\left(|\vec{x}-\vec{X}|\right). \notag
\end{eqnarray}
The  function $\psi_{n}\left(z\right)$ are the solutions of the eigenvalue equation 
\begin{eqnarray}
-h\left(z\right)^{-1}\partial_{z}\left(k\left(z\right)\partial_{z}\psi_{n}\right)=\lambda_{n}\psi_{n}\left(z\right),
\end{eqnarray}
with the eigenvalue  $\lambda_{n}$~\cite{Sakai:2004cn}, and
\begin{eqnarray}
\phi_{0}\left(z\right)
&=&\frac{1}{\sqrt{\kappa\pi}}\frac{1}{k\left(z\right)}, 
\nonumber \\
\phi_{n}\left(z\right) 
&=& \frac{1}{\sqrt{\lambda_{n}}}\partial_{z}\psi_{n}\left(z\right), \\
Y_{n}\left(r\right)
&=&-\frac{1}{4\pi}\frac{e^{-\sqrt{\lambda_{n}}}r}{r}, \ \ r=|\vec{x}|.
\end{eqnarray}

%=========================================
\subsection{\label{current}Currents}
%=========================================

Now we are ready to calculate the axial current. 
Following Ref.~\cite{Hashimoto:2008zw}, 
the chiral current is derived from the coupling with the external gauge field
$\delta\mathcal{A}_{\alpha}$
which is defined by
\begin{eqnarray}
\mathcal{A}_{\alpha}\left(x^{\mu},z\right)=\mathcal{A}^{cl}_{\alpha}\left(x^{\mu},z\right)
+\delta\mathcal{A}_{\alpha}\left(x^{\mu},z\right)
\end{eqnarray}
They are  related  to the left and right gauge fields in the four dimensional space at $z \to \pm \infty$, 
\begin{eqnarray}
\delta\mathcal{A}_{\mu}\left(x^{\nu},z\rightarrow+\infty\right)
=\mathcal{A}_{L\mu}\left(x^{\nu}\right), \notag \\
\delta\mathcal{A}_{\mu}\left(x^{\nu},z\rightarrow-\infty\right)
=\mathcal{A}_{R\mu}\left(x^{\nu}\right). \notag
\end{eqnarray}
Substituting this field into the action, the coefficients of the first order 
in $\mathcal{A}_{L\mu}, \ \mathcal{A}_{R\mu}$ is 
identified with the left and right currents $\mathcal{J}^{\mu}_{L}, \ \mathcal{J}^{\mu}_{R}$
with the sign properly taken into account, 
\begin{eqnarray}
&&\kappa\int d^{4}x\left[2{\rm tr}\left(\delta\mathcal{A}^{\mu}k\left(z\right)\mathcal{F}^{cl}_{\mu z}\right)\right]^{z=+\infty}_{z=-\infty}, \notag \\
&&=-2\int d^{4}x{\rm tr}\left(\mathcal{A}_{L\mu}\mathcal{J}^{\mu}_{L}+\mathcal{A}_{R\mu}\mathcal{J}^{\mu}_{R}\right) .
\end{eqnarray}
where
\begin{eqnarray}
&&\mathcal{J}^{\mu}_{L}=-\kappa\left(k\left(z\right)\mathcal{F}^{cl}_{\mu z}\right)\big|_{z=+\infty}, \notag \\
&&\mathcal{J}^{\mu}_{R}=+\kappa\left(k\left(z\right)\mathcal{F}^{cl}_{\mu z}\right)\big|_{z=-\infty}.
\end{eqnarray}
The vector and axial currents are then obtained by 
\begin{eqnarray}
\mathcal{J}^{\mu}_{V} 
&=& \mathcal{J}^{\mu}_{L} + \mathcal{J}^{\mu}_{R},
\notag \\
\mathcal{J}^{\mu}_{A}
&=&\mathcal{J}^{\mu}_{L} - \mathcal{J}^{\mu}_{R}
=
- \kappa\left[
\psi_{0}\left(z\right)k\left(z\right)\mathcal{F}^{cl}_{\mu z}
\right]^{z=+\infty}_{z=-\infty}, \label{axial current infty}
\end{eqnarray}
with $\psi_{0}\left(z\right)=\left(2/\pi\right)\arctan{z}$. 

When the instanton oscillates along the $z$ direction in a narrow range in the large $\lambda$ limit, 
the metrices are approximated as $h\left(Z\right)\simeq k\left(Z\right)\simeq1$. 
Then, 
substituting (\ref{instanton3}) for (\ref{axial current infty}) gives the following form
($r \equiv  |\vec x - \vec X|$)
\begin{eqnarray}
J^{i}_{A}(r; \vec X, Z, \rho, \vec a)
&=&
-2\pi^{2}\kappa\rho^{2}{\bm a}\tau^{a}{\bm a}^{-1}
\nonumber \\
& &
\hspace*{-1cm}
\times
\left(\left(\partial_{i}\partial_{a}-\delta^{ia}\partial^{2}_{j}\right)H^{A}-\epsilon^{iaj}\partial_{j}G^{A}\right)
\label{current position} 
\end{eqnarray}
where
\begin{eqnarray} 
G^{A}\left(r; \vec X, Z\right)=&&\left[\psi_{0}\left(z\right)k\left(z\right)\partial_{z}G\right]^{z=+\infty}_{z=-\infty} \notag \\
=&&-\sum^{\infty}_{n=1}g_{a^{n}}\psi_{2n}\left(Z\right)Y_{2n}\left(r\right), \\
H^{A}\left(r; \vec X, Z\right)=&&\left[\psi_{0}\left(z\right)k\left(z\right)H\right]^{z=+\infty}_{z=-\infty} \notag \\
=&&-\frac{1}{2\pi^{2}}\frac{1}{k\left(Z\right)}\frac{1}{r}-\sum^{\infty}_{n=1}\frac{g_{a^{n}}}{\lambda_{2n}}\partial_{Z}\psi_{2n}\left(Z\right)Y_{2n}\left(r\right), \notag \\ \\
g_{a^{n}}=&&\lambda_{2n}\kappa\int dzh\left(z\right)\psi_{2n}\psi_{0}.
\end{eqnarray}

To go further, it is convenient to present the Fourier transform in the momentum space, 
(in what follows the dependence  on the collective coordinates $\vec X, Z, \rho, \vec a$ are suppressed)
\begin{eqnarray}
\tilde{J}^{\mu}_{A}(\vec{q})
=
\int d^{3}x\ e^{-i\vec{q}\cdot\vec{x}}J^{\mu}_{A}(r). 
\end{eqnarray}
We obtain the following form:
%\begin{widetext}
\begin{eqnarray}
\tilde{J}^{cj}_{A}(\vec{q})
&=&
e^{-i\vec{q}\cdot\vec{X}}2\pi^{2}\kappa\rho^{2}{\rm tr}\left(\tau^{c}{\bm a}\tau^{a}{\bm a}^{-1}\right)
\nonumber \\
&\times&
\left(\delta_{aj}-\frac{q_{a}q_{j}}{\vec{q}^{2}}\right)\sum_{n\geq1}\frac{g_{a^{n}}\partial_{Z}\psi_{2n}\left(Z\right)}{\vec{q}^{2}+\lambda_{2n}}. \label{momentum axial current}
\end{eqnarray}
%\end{widetext}
This current is regarded as an operator in terms of the dynamical variable $\vec X$,  $Z$, $\rho$ and $\vec a$, 
which is used when taking the matrix elements by the corresponding wave functions.  

%=========================================
\section{\label{decay properties}Decay properties of Roper resonance}
%=========================================

Now, we investigate the decay properties of the Roper resonance, in particular 
the one pion emission decay $N^*(1440) \to \pi N$.  
Because the Roper resonance has a very large width causing uncertainties 
in the Breit-Wigner fitting, 
we refer to the result of the pole analysis.
Following the PDG table~\cite{Zyla:2020zbs}, we quote the following nominal values
\begin{eqnarray}
M_{N^*} &=& 1360 - 1380 \ (\sim 1370) \ {\rm MeV},
\\ \nonumber
\Gamma_{\rm total} &=& 160 -  190 \ (\sim 175) \ {\rm MeV},
\end{eqnarray}
and the branching ratio of the one pion decay
\begin{eqnarray}
N^* \to N \pi: \ \  55 - 75\  \%.
\end{eqnarray}
Using the lower and upper bounds for the total decay width and branching ratio, 
we find the partial decay width of the one pion decay
\begin{eqnarray}
\Gamma_{N^* \to \pi N} \sim 90 - 140\ {\rm MeV}.
\label{Gamma_exp}
\end{eqnarray}

%=========================================
\subsection{\label{axial coupling}Axial coupling $g_{A}$}
%=========================================

The axial coupling $g^{NN^{*}}_{A}$ for the transition $N^{*}\left(1440\right)\rightarrow N+\pi$ is defined as follows:
\begin{eqnarray}
\int d^{3}x\braket{N, s^{\prime}_{3} I^\prime_3 |J^{ai}_{A}|N^{*},s_{3}, I_3 }\times2
\nonumber \\
=
\frac{2}{3}g^{NN^{*}}_{A}\left(\sigma^{i}\right)_{s^{\prime}_{3},s_{3}}\left(\tau^{a}\right)_{I^{\prime}_{3},I_{3}}.
\end{eqnarray}
The factor 2/3 on the right hand side is needed in the chiral limit~\cite{Adkins:1983ya}.  
Using (\ref{momentum axial current}) and (\ref{wave func}), we obtain 
\begin{eqnarray}
g^{NN^{*}}_{A}\left(\vec{q}\right)=\frac{8\pi^{2}\kappa}{3}\braket{R_{N^{*}}|\rho^{2}|R_{N}}\sum_{n=1}\frac{g_{a_{n}}\braket{\partial_{Z}\psi_{2n}\left(Z\right)}}{\vec{q}^{2}+\lambda_{2n}} \notag \\
\end{eqnarray} 
where $\braket{\partial_{Z}\psi_{2n}\left(Z\right)}$ stands for the expectation value using the wave functions of $Z$.  
The matrix element of $\rho^2$ can be computed and the result is 
\begin{eqnarray}
\braket{R_{N^{*}}|\rho^{2}|R_{N}}&&=\left(1+2\sqrt{1+\frac{N^{2}_{c}}{5}}\right)^{-1/2}\braket{R_{N}|\rho^{2}|R_{N}} \notag \\
&&=\frac{\sqrt{5}}{2N_{c}}\left(1+2\sqrt{1+\frac{N^{2}_{c}}{5}}\right)^{1/2}\rho^{2}_{cl} \label{ratio}
\end{eqnarray}
with $\rho_{cl}$ being the classical instanton size given by
\begin{eqnarray}
\rho^{2}_{cl}&&=\frac{N_{c}}{8\pi^{2}\kappa}\sqrt{\frac{6}{5}}.
\end{eqnarray}
We note that the transition matrix element for $N^{*}\left(1440\right)\rightarrow N+\pi$
is related to the nucleon matrix element, an interesting feature 
of the present model associated with the collective nature of baryons.  
The axial coupling constant is then defined at $\vec q = 0$, $ g^{NN^{*}}_{A} = g^{NN^{*}}_{A}(\vec{0})$.
Using the relation
\begin{eqnarray}
\sum_{n=1}\frac{g_{a_{n}}\partial_{Z}\psi_{2n}\left(Z\right)}{\lambda_{2n}}=\frac{2}{\pi}\frac{1}{k\left(Z\right)},
\end{eqnarray} 
$g^{NN^{*}}_{a}$ can be expressed in a compact form: 
\begin{eqnarray}
g^{NN^{*}}_{A}=\frac{16\pi\kappa}{3}\braket{R_{N^{*}}|\rho^{2}|R_{N}}\Braket{\frac{1}{k\left(Z\right)}}.
\end{eqnarray}
In the above equations, $\langle \cdots \rangle$ stands for the expectation value using the wave functions of $Z$.

There are two parameters of this model, $M_{KK}$ and $\kappa$.
Following  Adkins et al~\cite{Adkins:1983ya} they are determined to reproduce 
the mass splitting of the nucleon and delta, and the pion decay constant 
$f_\pi = 64.5$ MeV, 
\begin{eqnarray}
M_{KK}=488 \ {\rm MeV}, \ \ \ \kappa=0.0137.
\end{eqnarray}
Then, the prediction of the present model for $g^{NN^{*}}_{A}$ is
\begin{eqnarray}
g^{NN^{*}}_{A}=0.402. \label{numerical gA}
\end{eqnarray}

%=========================================
\subsection{\label{decay width}Decay width}
%=========================================

The decay width of $N^{*}\left(1440\right)\rightarrow N+\pi$ can be computed by the formula
\begin{eqnarray}
\Gamma_{N^{*}\left(1440\right)\rightarrow N+\pi}
&=&
\frac{1}{2M_{N^{*}}} 
\int\frac{d^{3}p_{N}}{\left(2\pi\right)^{3}2E_{N}}\frac{d^{3}p_{\pi}}{\left(2\pi\right)^{3}2E_{\pi}}
\nonumber \\
&\times&
\left(2\pi\right)^{4}\delta^{4}\left(p_{N}+p_{\pi}\right)\left|t_{fi}\right|^{2}, \label{decay width def}
\end{eqnarray}
where the amplitude $t_{fi}$ is given by the Lagrangian 
\begin{eqnarray}
L=i\frac{M_{N}+M_{N^{*}}}{2f_{\pi}}g^{NN^{*}}_{A}\overline{\psi}_{N^{*}}\gamma_{5}\vec{\tau}\cdot\vec{\pi}\psi_{N}+h.c.,
\end{eqnarray}
as follows
\begin{eqnarray}
t_{fi}
&=&\braket{N(-\vec{q})\pi(\vec{q})|L|N^{*}(\vec{0})} \notag \\
&=&\sqrt{2M_{N^{*}}}\sqrt{M_{N}+E_{N}} \nonumber \\
&\times&
\frac{M_{N}+M_{N^{*}}}{2f_{\pi}}\frac{g^{NN^{*}}_{A}}{E_{N}+M_{N}}\braket{s^{\prime}_{3}|\vec{\sigma}\cdot\vec{q}|s_{3}}.
\end{eqnarray}
Here we have expressed the effective $\pi N N^*$ coupling $g_{\pi NN^*}$ 
in terms of the axial coupling by using the Goldberger-Treiman relation,
\begin{eqnarray}
g^{NN^{*}}_{A}=\frac{f_{\pi}g_{\pi NN^{*}}}{\left(M_{N}+M_{N^{*}}\right)/2} .
\end{eqnarray}
Hence we obtain
\begin{eqnarray}
&\Gamma&\! \! _{N^{*}\left(1440\right)\rightarrow N+\pi}
\nonumber \\
&=&\frac{q}{4\pi}\frac{M_{N}+E_{N}}{M_{N^{*}}}
\left(\frac{M_{N}+M_{N^{*}}}{2f_{\pi}}\frac{g^{NN^{*}}_{A}q}{E_{N}+M_{N}}\right)^{2}. 
\label{decay width NN*}
\end{eqnarray}
Using 
$
M_{N}=940 \ {\rm MeV}, \ M_{N^{*}}=1370 \ {\rm MeV},  \ q=342 \ {\rm MeV}, 
$
we find
\begin{eqnarray}
\Gamma_{N^{*}\left(1440\right)\rightarrow N+\pi}= 64 \ {\rm MeV}.
\end{eqnarray}
In this computation the value of $g^{NN^{*}}_{A}$ at $\vec q = 0$ is used.   
By considering the form factor effect, the $g^{NN^{*}}_{A}$ 
value at $\vec q = 342$ MeV becomes about 15 \% smaller, and hence 
$\Gamma_{N^{*}\left(1440\right)\rightarrow N+\pi} \sim 55$ MeV.  

These values are smaller than the experimental value (\ref{Gamma_exp}). 
This is because the axial coupling $g^{NN^{*}}_{A}$ is small, 
which is a common feature of the solitonic picture of baryons. 
In fact, the nucleon $g_A^{NN}$ is computed in a similar manner as for $g_A^{NN^*}$ by 
using the nucleon wave function $R_N(\rho)$.
The result is 
\begin{eqnarray}
g_A^{NN} = 0.837.
\end{eqnarray}
Though small, it is interesting to observe that the value is somewhat larger than that of~\cite{Adkins:1983ya}.  
One possible resolution to recover the experimental value 
$g_A^{NN} = 1.25$
is to take into account $1/N_c$ corrections (Ref.~\cite{Hosaka:1996ee} and 
references there).  
Here, however, we do not discuss this anymore.  
On the other hand, it is interesting 
to observe the relation between the axial couplings of the nucleon and that of 
the Roper-nucleon transition.  
Inspection of Eq. (\ref{ratio}), we find
\begin{eqnarray}
g^{NN^{*}}_{A} : g^{NN}_{A}
&=&
1:\left(1+2\sqrt{1+\frac{N^{2}_{c}}{5}}\right)^{1/2}
\nonumber \\
&=&1:2.08. 
\label{gA_ratio}
\end{eqnarray}
We emphasize that this relation does not include any model parameters (except for 
$N_c = 3$), and so 
a model independent relation.  
Experimentally, if we use the partial decay width $\Gamma_{N^* \to \pi N} \sim 110$ MeV, 
we find the ratio
\begin{eqnarray}
g^{NN^{*}}_{A}:g^{NN}_{A} = 0.77:1.25 \sim 1:1.62,
\end{eqnarray}
which agrees well with the present model prediction within $\sim$ 20 \% accuracy.  

%==========================================
\section{\label{conclusion}Discussions and summary}
%==========================================

In this paper, we have studied one pion emission decay of the Roper resonance, 
$N^{*}\left(1440\right)\rightarrow N+\pi$ in the Sakai-Sugimoto model of the Holographic QCD.
Baryons are described as collective states of instantons of the five-dimensional Yang-Mills theory.
We have then employed the currents as defined in Ref.~\cite{Hashimoto:2008zw}, 
and computed the matrix elements.  
Resulting axial coupling has turned out to be too small as compared to 
what is expected from the experimental data.
This is a rather common feature of the solitonic model for baryons.  
However, an important finding has been made for the ratio of $g_A$'s of the nucleon and 
Roper-nucleon transition in a model independent manner.  

The present picture of baryons  as instantons with collective dynamics 
is very much the same as the Skyrmion picture, baryons as chiral solitons.
In contrast, 
it is very much different from the conventional quark model one, where 
baryons are described by single particle states of the constituent quarks.  
As anticipated, the quark model gave only a tiny decay rate for the Roper resonance
when the leading term in $1/m$ expansion of the quark-pion interaction is used,
which has been the widely adopted prescription.  
For this problem a resolution has been recently proposed by including higher order terms of $1/m^2$~\cite{Arifi:2021orx}.
In the quark model, however, the model independent relation between $g_A^{NN^*}$ and $g_A^{NN}$
is not derived.  
In this respect,  such model independent relations would be helpful to further 
investigate the nature of nucleon resonances.  

\begin{acknowledgments}
This work is supported in part by JSPS KAKENHI No. JP17K05441 (C) 
and Grants-in-Aid for Scientific Research on Innovative Areas (No. 18H05407). 
\end{acknowledgments}

\appendix

\bibliography{roperdecay}% Produces the bibliography via BibTeX.

\end{document}